\documentclass[aps,amsfonts,twocolumn,nofootinbib,floatfix]{revtex4}
\usepackage{amsmath}
\usepackage{epsfig}
\usepackage{latexsym}
\usepackage{color}
\usepackage{pstricks}

\begin{document}

\title{A late-accelerating universe with no dark energy\\
 -- and a finite-temperature big bang}

\author{Richard A. Brown$^{1}$, Roy Maartens$^{1}$,
Eleftherios Papantonopoulos$^{2}$, Vassilis Zamarias$^{2}$}

\affiliation{\vspace*{0.2cm} $^1$Institute of Cosmology \&
Gravitation, University of Portsmouth, Portsmouth~PO1~2EG, UK}

\affiliation{ $^2$Department of Physics, National Technical
University of Athens, GR~157~73~Athens, Greece}

\date{\today}

\begin{abstract}

Brane-world models offer the possibility of explaining the late
acceleration of the universe via infra-red modifications to
General Relativity, rather than a dark energy field. However, one
also expects ultra-violet modifications to General Relativity,
when high-energy stringy effects in the early universe begin to
grow. We generalize the DGP brane-world model via an ultra-violet
modification, in the form of a Gauss-Bonnet term in the bulk
action. The combination of infra-red and ultra-violet
modifications produces an intriguing cosmology. The DGP feature of
late-time acceleration without dark energy is preserved, but there
is an entirely new feature -- there is no infinite-temperature big
bang in the early universe. The universe starts with finite
density and pressure, from a ``sudden" curvature singularity.

\end{abstract}

\maketitle

\section{Introduction}

The standard cosmology based on General Relativity and inflation
has been remarkably successful. But there remain deep puzzles left
for theorists to resolve -- what is the cause of the late-time
acceleration of the universe (the ``dark energy" problem)? how is
the classical big bang singularity removed by quantum gravity
effects? One approach to start tackling these problems is via the
brane-world scenario, which is motivated by string theory. Most
brane-world models, including those of Randall-Sundrum
type~\cite{rs}, produce ultra-violet modifications to General
Relativity, with extra-dimensional gravity dominating at high
energies. However it is also possible for extra-dimensional
gravity to dominate at low energies, leading to infra-red
modifications of General
Relativity~\cite{Gregory:2000jc,Dvali:2000hr}. The
Dvali-Gabadadze-Porrati (DGP) models~\cite{Dvali:2000hr} (see
also~\cite{dgp0}) achieve this via a brane induced gravity effect.

The generalization of the DGP models to cosmology lead to
late-accelerating cosmologies, even in the absence of a dark
energy field~\cite{Deffayet:2000uy}. This exciting feature of
``self-acceleration" may help towards a new resolution to the dark
energy problem. But the models suffer from the short-coming that
they do not modify 4D gravity at high energies, where we expect
stringy corrections to start having an effect. How can we
generalize the DGP models so that they also show ultra-violet
modifications to General Relativity? One possibility is to
introduce into the gravitational action a term that is associated
with higher-energy stringy corrections -- the Gauss-Bonnet
term~\cite{gb}. Indeed, in certain realizations of string theory,
the ghost-free GB term in the bulk action may naturally lead to a
DGP induced gravity term on the brane
boundary~\cite{Mavromatos:2005yh}.

We investigate what happens when a Gauss-Bonnet term is introduced
in the 5D Minkowski bulk containing a Friedmann brane with DGP
induced gravity. As we will show, this combination of infra-red
and ultra-violet modifications leads to intriguing cosmological
models.

\section{Field equations}

The gravitational action contains the Gauss-Bonnet (GB) term in
the bulk, as a correction to the Einstein-Hilbert term, and the
Induced Gravity (IG) term on the brane,
\begin{eqnarray}\label{AcGBIG}
S_{\rm grav}&=&\frac{1}{2\kappa_5^2}\int d^5 x\sqrt{-
g^{(5)}}\left\{ R^{(5)} \right.\nonumber\\&+&\left.\alpha\left[
R^{(5)2}-4 R^{(5)}_{ab}R^{(5)ab}+
R^{(5)}_{abcd}R^{(5)abcd}\right]\right\}
\nonumber\\&+&\frac{r}{2\kappa^2_5} \int_{\rm brane}d^4x\sqrt{-
g^{(4)}}\,R^{(4)}\,,
\end{eqnarray}
where $\alpha\,(\geq0)$ is the GB coupling constant~\cite{alpha}
and $r\, (\geq 0)$ is the IG ``cross-over" scale, which marks the
transition from 4D to 5D gravity. The DGP models are the special
case $\alpha=0$, and in this case the cross-over scale defines an
effective 4D gravitational constant via
${\kappa^2_4}={\kappa^2_5}/r\,. $

We assume mirror ($Z_2$) symmetry about the brane. The standard
energy conservation equation holds on the brane,
\begin{equation}\label{ec}
\dot \rho+3H(1+w)\rho=0\,,~w=p/\rho\,.
\end{equation}
The modified Friedmann equation was found in the most general case
(where the bulk contains a black hole and a cosmological constant,
and the brane has tension) in Ref.~\cite{Kofinas:2003rz}. For a
spatially flat brane without tension, in a Minkowski bulk,
\begin{eqnarray}\label{F}
4\left[1+\frac{8}{3}
\alpha\left(H^2+\frac{\Phi}{2}
\right)\right]^2 \left(H^2-\Phi \right)
=\left[rH^2 
-\frac{\kappa^2_5}{3}\rho \right]^2\!,
\end{eqnarray}
where $\Phi$ is determined by
\begin{equation}\label{Phi}
\Phi+2\alpha\Phi^2=0\,.
\end{equation}
(In the most general case, the right-hand side of this condition
is nonzero~\cite{Kofinas:2003rz}.) Equation~(\ref{Phi}) has
solutions $\Phi=0, -1/2\alpha$, but here we only consider
$\Phi=0$, since the second solution has no IG limit and thus does
not include the DGP model~\cite{ads}.

\begin{figure}
\includegraphics[height=3in,width=2.75in,angle=270]{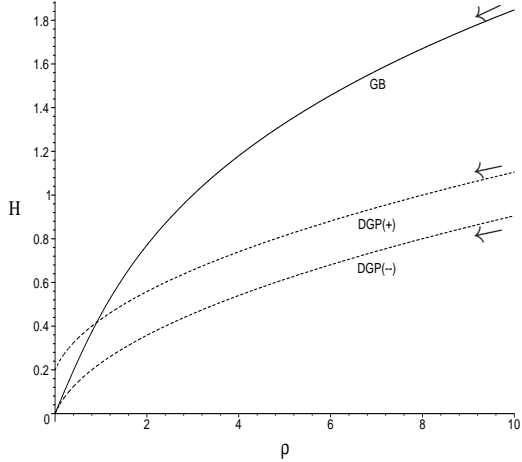}
\rput{25}(-1.2,-0.85){\large $\leftarrow$}
\rput{12}(-1.2,-2.95){\large $\leftarrow$}
\rput{12}(-1.2,-3.8){\large $\leftarrow$} \caption{DGP and GB
solutions of the Friedmann equation ($H$ vs $\rho$) for a
Minkowski bulk. (Arrows indicate the flow of brane proper time
$t$, with $t=\infty$ at $\rho=0$.)} \label{GBDGP}
\end{figure}

The DGP models have $\alpha=\Phi=0$ and the Friedmann
equation~(\ref{F}) reduces to a quadratic in $H^2$, with solutions
\begin{equation}\label{FD}
H^2=\pm {2\over r}H+{\kappa_5^2 \over 3r}\rho\,.
\end{equation}
There are two branches DGP($\pm$) for the two signs on the right
(corresponding to different embeddings of the brane in the
Minkowski bulk). Both branches have a 4D limit at high energies,
\begin{equation}\label{bbdgp}
\text{DGP($\pm$):}~~H\gg r^{-1}~\Rightarrow ~ H^2\propto \rho\,,
\end{equation}
while at low energies,
\begin{eqnarray}
\text{DGP(+):} && \rho \to 0 ~\Rightarrow ~ H\to {2 \over r}\,,\\
\text{DGP(--):} && \rho \to 0 ~\Rightarrow ~ H^2\propto \rho^2\,.
\end{eqnarray}
DGP(--) has a non-standard (and non-accelerating) late universe.
The self-accelerating DGP(+) branch is of most interest for
cosmology, and we focus here on this model and its generalization
via a GB term.

The pure GB model with a Minkowski bulk has $r=\Phi=0$ and the
Friedmann equation~(\ref{F}) reduces to
\begin{equation}\label{FG}
\left(1+\frac{8}{3} \alpha H^2\right)^2H^2={\kappa_5^4 \over
36}\rho^2\,,
\end{equation}
which is a cubic in $H^2$. This GB Friedmann equation has no 4D
limit:
\begin{eqnarray}
\text{GB high energy:} && H\gg \alpha^{-1/2} ~\Rightarrow ~
H^2\propto \rho^{2/3} \,,\label{bbgb}\\
\text{GB low energy:} && H\ll \alpha^{-1/2} ~\Rightarrow ~
H^2\propto \rho^2\,.
\end{eqnarray}

The Friedmann equations for pure DGP and pure GB models with a
Minkowski bulk are compared in Fig.~\ref{GBDGP}.

\section{DGP brane with GB bulk gravity:
combining UV and IR modifications}

The DGP(+) models are attractive for cosmology since they
accelerate at late times, without the need for dark energy, when
gravity begins to leak off the brane, i.e., when the 5D Ricci term
in Eq.~(\ref{AcGBIG}) begins to dominate over the 4D Ricci term.
At early times, the 4D term dominates and General Relativity is
recovered (in the background -- note that the perturbations are
not General Relativistic~\cite{dgppert}). The DGP models are in
some sense ``unbalanced", since they do not include ultra-violet
modifications to cosmological dynamics. In order to modify 4D
gravity at high energies as well as low energies, we can include a
GB term in the action.

The combined Gauss-Bonnet Induced Gravity (GBIG) model has a DGP
brane in a Minkowski bulk with Einstein-Gauss-Bonnet gravity. The
GBIG Friedmann equation follows from putting $\Phi=0$ in
Eq.~(\ref{F}). Defining dimensionless variables,
\begin{eqnarray}\label{DV}
\gamma&=&\frac{8\alpha}{3r^2}\,,~ h=Hr\,,~ \mu
=\frac{r\kappa^2_5}{3}\rho\,,~\tau={t \over r}\,,
\end{eqnarray}
the GBIG Friedmann equation becomes
\begin{equation}\label{DGPF2}
4\left(\gamma h^2+1\right)^2h^2=\left(h^2-\mu\right)^2,
\end{equation}
while the conservation equation becomes
\begin{equation}\label{ec2}
\mu'+3h(1+w)\mu=0\,,
\end{equation}
where a dash denotes $d/d\tau$, and $h=a'/a$.

Combining Eqs.~(\ref{DGPF2}) and (\ref{ec2}), we find the modified
Raychaudhuri equation,
\begin{equation}\label{Ray}
{h}'=\frac{3\mu(1+w)(h^2-\mu)}{4(\gamma h^2+1)(3\gamma
h^2+1)-2(h^2-\mu)}\,.
\end{equation}
The acceleration $a''/a=h'+h^2$ is then given by
\begin{equation}\label{acc}
{a'' \over a}={4h^2(\gamma h^2+1)(3\gamma h^2+1)-(h^2-\mu)
[2h^2-3(1+w)\mu] \over 4(\gamma h^2+1)(3\gamma h^2+1)-2(h^2-\mu)}.
\end{equation}

\begin{figure}
\begin{center}
\includegraphics[height=3in,width=2.75in,angle=270]{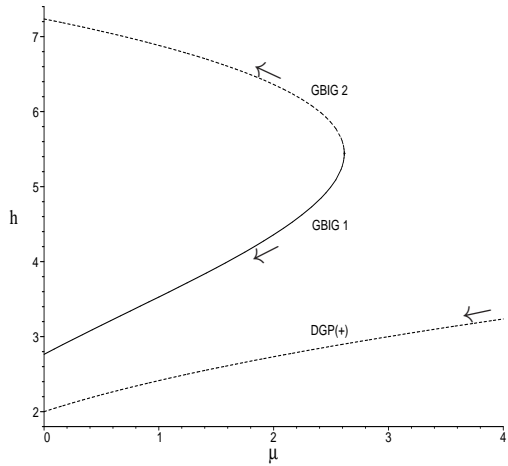}
\rput{335}(-4,-1.6){\large $\leftarrow$} \rput{27}(-4,-4){\large
$\leftarrow$} \rput{12}(-1.2,-4.8){\large $\leftarrow$}
\caption{Solutions of the Friedmann equation ($h$ vs $\mu$) for
the DGP(+) model and its Gauss-Bonnet corrections, GBIG1 and
GBIG2. The curves are independent of the equation of state $w$.
Arrows indicate the flow of normalized brane proper time $\tau$,
with $\tau=\infty$ at $\mu=0$. (Here $\gamma=0.05$.)}
\label{GBIGDGP}
\end{center}
\end{figure}

The GB correction, via a non-zero value of $\gamma$, introduces
significant complexity to the Friedmann equation, which becomes
cubic in $h^2$, as opposed to the quadratic DGP($\pm$) case,
$\gamma= 0$, for which
\begin{equation}\label{dd}
h^2=\mu+2\pm2\sqrt{\mu+1}\,.
\end{equation}
This additional complexity has a dramatic effect on the dynamics
of the DGP(+) model, as shown in Fig.~\ref{GBIGDGP}. The
contribution of GB gravity at early times removes the infinite
density big bang, and the universe starts at finite maximum
density and finite pressure (but, as we show below, with infinite
curvature). Furthermore, there are two such solutions, each with
late-time self-acceleration, marked GBIG1 and 2 on the plots.
Since GBIG2 is accelerating throughout its evolution (actually
super-inflating, $h'>0$), the physically relevant
self-accelerating solution is GBIG1.

\begin{figure}
\begin{center}
\includegraphics[height=3in,width=2.75in,angle=270]{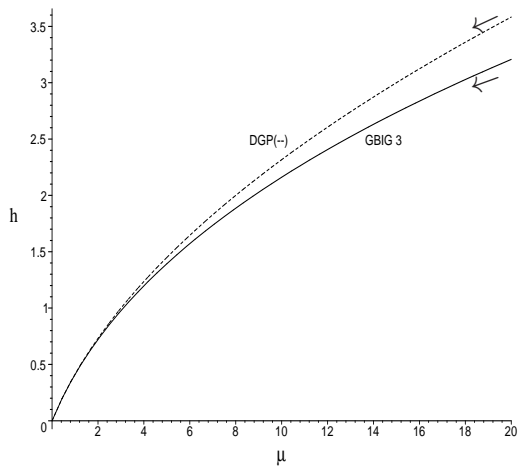}
\rput{25}(-1.2,-0.9){\large $\leftarrow$}
\rput{20}(-1.2,-1.7){\large $\leftarrow$} \caption{The DGP(--)
model and its GB correction, GBIG3. } \label{GBIGDGP2}
\end{center}
\end{figure}

The cubic in $h^2$, Eq.~(\ref{DGPF2}), has three real roots when
$0<\gamma<1/16$ (see below). Two of these roots correspond to
GBIG1--2, which are modifications of the DGP(+) model. The third
root GBIG3 is a modification of the DGP(--) model, as illustrated
in Fig.~\ref{GBIGDGP2}. Note that the curves in these figures are
independent of the equation of state $w$ of the matter content of
the universe -- $w$ will determine the time evolution of the
universe along the curves, via the conservation
equation~(\ref{ec2}).

The plots show that GBIG3 starts with a standard big bang,
$\rho=\infty$, in common with the DGP($\pm$) and GB models in
Fig.~\ref{GBDGP}. By contrast, GBIG1--2 have a finite-temperature
big bang, since the density is bounded above,
\begin{equation}
\mu \leq \mu_{\rm i}\,,
\end{equation}
where $\mu_{\rm i}$ (which is positive only for $\gamma<1/16$), is
found below, in Eq.~(\ref{rhomax}).

The finite-density beginning was pointed out in
Ref.~\cite{Kofinas:2003rz}, where the cubic for the general case
(i.e., with tension, bulk cosmological constant and bulk black
hole) was qualitatively analyzed. Here, we focus on the simplest
generalization of the DGP models, and give a detailed quantitative
analysis of the cosmological dynamics. In particular, our analysis
shows that one solution is not bounded, which was not noticed in
Ref.~\cite{Kofinas:2003rz}. The numerical plots of the Friedmann
equation in Figs.~\ref{GBIGDGP} and \ref{GBIGDGP2} are crucial to
a proper understanding of the algebraic analysis of the cubic
roots.

A detailed analysis~\cite{bmp} of the cubic equation~(\ref{DGPF2})
confirms the numerical results, and shows that (for $\mu>0$):
\begin{eqnarray}
0<\gamma<{1\over 16} &:& \mbox{3 real roots, GBIG1--3,}
\label{class}
\\ \gamma\geq {1\over 16} &:& \mbox{1 real root,~ GBIG3.}
\end{eqnarray}
The real roots are given as follows:\\

\noindent$\bullet$ For $0<\gamma<1/16$: the roots GBIG1--2 are
\begin{equation}\label{roots1}
4\gamma^2 h^2= {1-8\gamma \over 3}+ 2\sqrt{-Q} \cos(\theta+{n\pi
\over 3}) ~\mbox{for}~\mu \leq \mu_{\rm i}\,,
\end{equation}
where $n=4$ for GBIG1, $n=2$ for GBIG2, and the root GBIG3 is
\begin{equation}\label{roots2}
4\gamma^2 h^2= {1-8\gamma \over 3}+\left\{ \begin{array}{ll}
2\sqrt{-Q} \cos\theta & \mbox{for}~\mu \leq \mu_{\rm i},
\\ \\ S_++S_- & \mbox{for}~\mu\geq \mu_{\rm i}\,.
\end{array}\right.
\end{equation}
$\bullet$ For $\gamma\geq 1/16$: the only real root GBIG3 is
\begin{equation}\label{roots3}
4\gamma^2 h^2= {1-8\gamma \over 3}+ S_++S_- \,.
\end{equation}\\

In the above, $S_\pm, Q,R,\theta$ are defined by
\begin{eqnarray}
&&S_\pm = \left[R\pm \sqrt{R^2+Q^3}\right]^{1/3},\\
&& Q = {8\gamma^2 \over 3}(\mu+2)-{1\over 9}(1-8\gamma)^2\,,\\
&&R = 8\gamma^4\mu^2- {4\gamma^2 \over 3}(1-8\gamma)(\mu+2)
+{(1-8\gamma)^3\over 27}\!, \label{R} \\ && \cos 3\theta = R/
\sqrt{-Q^{3}}\,. \label{theta}
\end{eqnarray}

The explicit form of the solutions can be used to confirm the
features in Figs.~\ref{GBIGDGP} and \ref{GBIGDGP2}.
Equations~(\ref{roots2})--(\ref{R}) show that GBIG3 starts with a
big bang, $h,\mu\to\infty$, with $h^2 \sim \mu^{2/3}$ near the big
bang. This is the same as the high-energy behaviour of the pure GB
model, Eq.~(\ref{bbgb}) -- the GB effect dominates at high
energies in GBIG3. This is not the case for GBIG1--2.

The maximum density feature of GBIG1--2 is more easily confirmed
by analysing the turning points of $\mu$ as a function of $h^2$.
The Friedmann equation~(\ref{DGPF2}) gives
\begin{equation}\label{in}
{d\mu \over d(h^2)}={h^2-\mu-2(\gamma h^2+1)(3\gamma h^2+1)\over
h^2-\mu}\,.
\end{equation}
Substituting $d\mu/d(h^2)=0$ into Eq.~(\ref{DGPF2}), we find that
\begin{eqnarray}
h_{\rm i}&=&\frac{1\pm\sqrt{1-12\gamma}}{6\gamma}\,,
\\
\mu_{\rm i} &=& \pm{1\over 3} h_{\rm
i}^2\left(2\sqrt{1-12\gamma}\mp 1 \right)\,.
\end{eqnarray}
The second equation shows that positive maximum density only
arises for the upper sign and with $\gamma<1/16$, in agreement
with the cubic analysis. Thus the initial Hubble rate and density
for GBIG1--2 are
\begin{eqnarray}
h_{\rm i}&=&\frac{1+\sqrt{1-12\gamma}}{6\gamma}\,, \label{Hrhomax}
\\
\mu_{\rm i} &=& {1\over 3} h_{\rm i}^2\left(2\sqrt{1-12\gamma}- 1
\right)\,. \label{rhomax}
\end{eqnarray}
If $\gamma=0$, then GBIG1--2 reduce to DGP(+), and $h_{\rm
i}=\mu_{\rm i}=\infty$. Note that
\begin{equation}
h_{\rm i}>4\,.
\end{equation}
The case $\gamma=1/16, \mu_{\rm_i}=0$ corresponds to a vacuum
brane with de Sitter expansion, and $h=h_{\rm i}=4$, generalizing
the DGP(+) vacuum de Sitter solution~\cite{Deffayet:2000uy}.

The late-time asymptotic value of the expansion rate, as $\mu\to
0$, is
\begin{equation}\label{GBIGhinfin}
h_\infty={1\over 2\sqrt{2}\,\gamma}\left[1-8\gamma\mp
\sqrt{1-16\gamma}\right]^{1/2},
\end{equation}
where the minus sign corresponds to GBIG1 and the plus sign to
GBIG2. In the limit $\gamma\to 0$, GBIG1 recovers the DGP(+) case,
$h_\infty=2$, while for GBIG2, $h_\infty\to\infty$; the parabolic
GBIG1--2 curve in Fig.~\ref{GBIGDGP} ``unwraps" and transforms
into the DGP(+) curve. Equations~(\ref{class}) and
(\ref{GBIGhinfin}) show that
\begin{equation}\label{cross}
2\leq h_\infty <4 ~~\mbox{for GBIG1,}
\end{equation}
while $4<h_\infty<\infty$ for GBIG2.

\begin{figure}
\begin{center}
\includegraphics[height=3in,width=2.75in,angle=270]{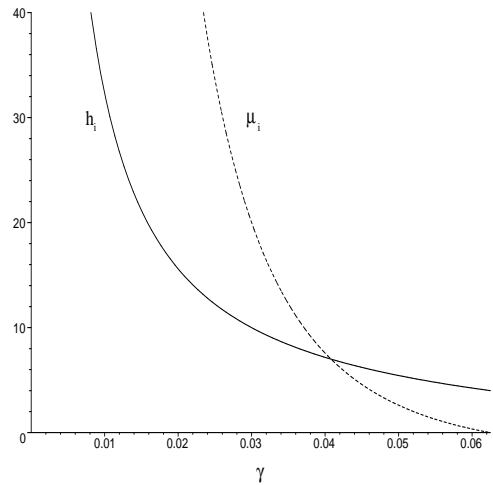}
\caption{The dependence in GBIG1--2 of the initial expansion rate
and density on $\gamma$. } \label{himui2}
\end{center}
\end{figure}

\begin{figure}
\begin{center}
\includegraphics[height=3in,width=2.75in,angle=270]{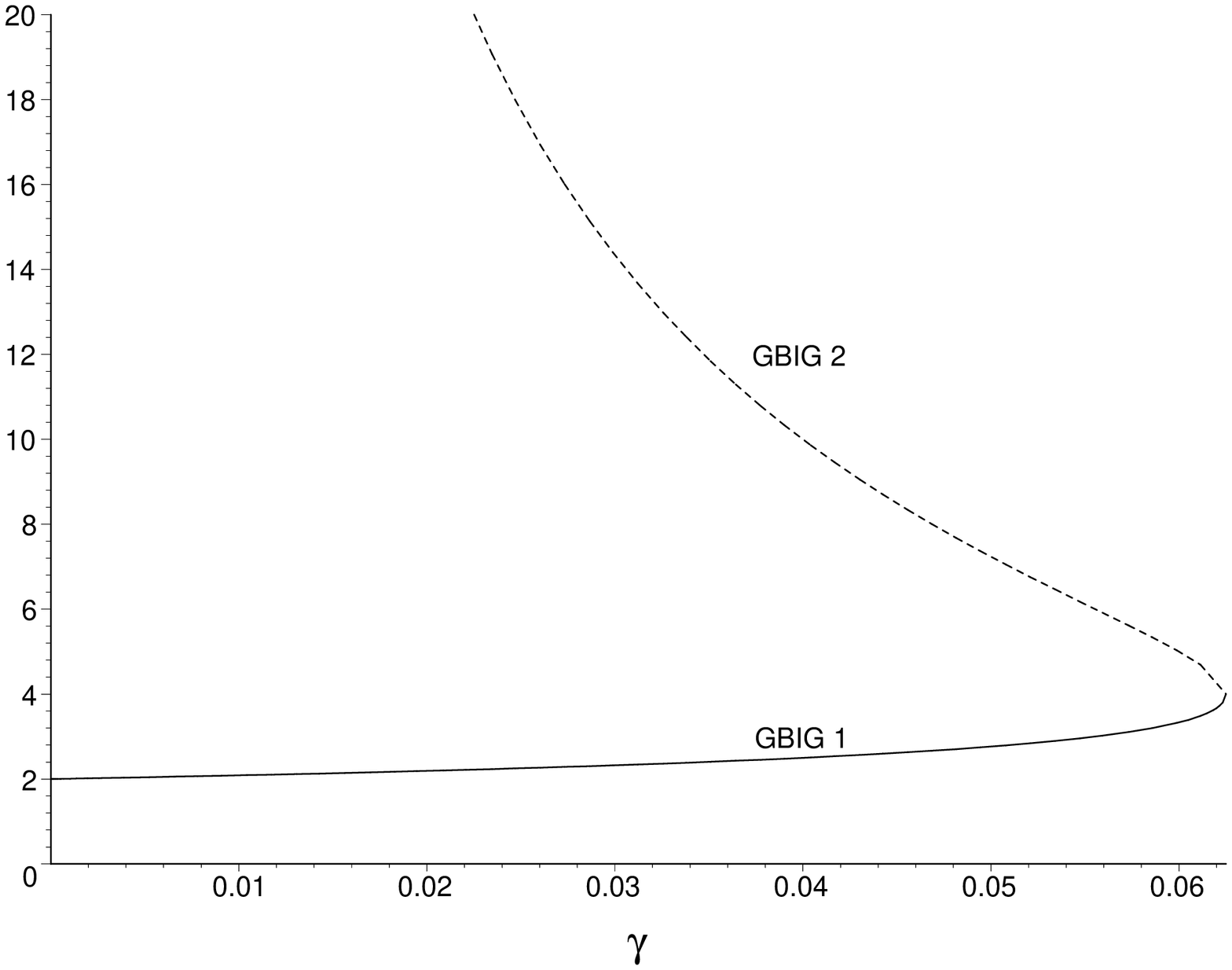}
\rput{0}(-7.4,-3.2){\large $h_{\infty}$}
\caption{The GBIG1--2 late-time asymptotic expansion rate as a
function of $\gamma$. } \label{hinfin}
\end{center}
\end{figure}

The behaviour of the key GBIG1--2 parameters is illustrated in
Figs.~\ref{himui2} and \ref{hinfin}.

\section{Cosmological Dynamics}

The GBIG1 model, which is the physically relevant generalization
of the DGP(+) model, exists if Eq.~(\ref{class}) holds. By
Eq.~(\ref{DV}), this means that the GB length scale $L_{\rm
gb}=\sqrt{\alpha}$ must be below a maximum threshold determined by
the IG cross-over scale:
\begin{equation}\label{gam2}
\gamma<{1\over 16} \,\Leftrightarrow\, L_{\rm gb}\equiv \sqrt
\alpha < {1\over 8}\sqrt{3 \over 2}\,r\,.
\end{equation}
If the GB term is taken as the correction term in certain string
theories, then $L_{\rm gb}\sim L_{\rm string}$, while $r\sim
H_0^{-1}$, so that this bound is easily satisfied.

When Eq.~(\ref{gam2}) holds, the universe starts with a maximum
density $\rho_{\rm i}$ and maximum Hubble rate $H_{\rm i}$, and
evolves to an asymptotic vacuum de Sitter state:
\begin{eqnarray}
0<\rho &<& \rho_{\rm i}= {rH_{\rm i}^2 \over
\kappa_5^2}\left(2\sqrt{1-{32\alpha \over r^2}}-1 \right),\\
H_\infty< H &<& H_{\rm i}= {r \over 16\alpha}\left(
\sqrt{1-{32\alpha \over r^2}}+1\right).
\end{eqnarray}
At any epoch $\tau_0$, the proper time back to the beginning is
\begin{equation}
\tau_0-\tau_{\rm i}=\int_{a_{\rm i}}^{a_0}\,{da \over ah}\,.
\end{equation}
Since $a$ and $h$ are nonzero on the interval of integration, the
time back to the beginning is finite.

The current Hubble rate can be approximated by the final de Sitter
Hubble rate, $H_0\sim H_\infty$, so that by Eq.~(\ref{cross}), the
cross-over scale obeys
\begin{equation}\label{hub}
2H_0^{-1} \lesssim r \lesssim 4H_0^{-1}\,.
\end{equation}
In the DGP(+) limit, $r\sim 2H_0^{-1}$. The effect of GB gravity
is to increase $r$ but not beyond $r\sim 4H_0^{-1}$.

However, there is a UV-IR ``bootstrap" operating to severely limit
the GB effect at late times. The key point is that appreciable
late-time GB effects require an increase in $\gamma$, whereas the
primordial Hubble rate $H_{\rm i}$ is suppressed by an increase in
$\gamma$ -- as shown in Figs.~\ref{himui2} and \ref{hinfin}.
Equations~(\ref{Hrhomax}) and (\ref{GBIGhinfin}) imply that
\begin{equation}\label{boot}
H_{\rm i}\gg H_0~ \Rightarrow~ \gamma \ll {1\over 16}\,.
\end{equation}
Thus the GBIG1 model does not alleviate the DGP(+) fine-tuning
problem of a very large cross-over scale, $r\sim H_0^{-1} \sim
(10^{-33}\,\mbox{eV})^{-1}$.

The GBIG1 Friedman equation~(\ref{roots1}) gives
\begin{eqnarray}
\!\!&& H^2={1-8\gamma\over 12\gamma^2r^2}\nonumber \\ \!\! &&+
{\sqrt{8\gamma^2(r \kappa_5^2 \rho + 6 ) -(1-8\gamma)^2}
 \over 6\gamma^2r^2}\cos\left[\theta(\rho) +{4\pi \over 3} \right]
\label{root}
\end{eqnarray}
where
\begin{eqnarray}
&&\cos3 \theta(\rho) = \nonumber \\ && ~~~{216 \gamma^4 \mu^2
-36\gamma^2(1-8\gamma)(\mu+2) +(1-8 \gamma)^3 \over [(1-8\gamma)^2
-24\gamma^2(\mu+2)]^{3/2} }.
\end{eqnarray}
A more convenient form of the Friedmann equation follows from
solving Eq.~(\ref{DGPF2}) for $\mu$,
\begin{equation}\label{fnew}
\mu=h^2-2h(\gamma h^2+1)\,,~~ h_\infty \leq h < h_{\rm i}\,.
\end{equation}
By expanding to first order in $h^2-h_\infty^2$, we find that at
late times,
\begin{equation}\label{late}
h^2 = h_\infty^2+2\left({h_\infty \over 4 -h_\infty}\right) \mu
+O(\mu^2)\,.
\end{equation}
Taking the DGP(+) limit $h_\infty\to 2$, and comparing with
Eqs.~(\ref{FD}) and (\ref{dd}), we find that the effective Newton
constant in GBIG1 is
\begin{equation}\label{newt}
G=\left({h_\infty \over 4 -h_\infty}\right){G_5 \over r} \,,
\end{equation}
where $G_5=\kappa_5^2/8\pi$ is the fundamental, 5D gravitational
constant. In the DGP(+) case $G=G_5/r$.

Equation~(\ref{newt}) gives a relation for the fundamental Planck
scale $M_5$
\begin{equation}\label{newt2}
M_5^3\sim \left({rH_0 \over 4 -rH_0}\right){M_{\rm p}^2 \over r}
\,,
\end{equation}
where $M_{\rm p}$ is the effective 4D Planck scale, and we used
$H_\infty \sim H_0$. As $r\to 4H_0^{-1}$ (its upper limit), so
$M_5$ increases. This is very different from the DGP(+) case,
where $M_5^3= M_{\rm p}^2/r$, so that $M_5$ is constrained to be
very low, $M_5 \lesssim 100\,$MeV. In principle, GB gravity allows
us to solve the problem of a very low fundamental Planck scale --
but in practice the UV-IR bootstrap, Eq.~(\ref{boot}), means that
$\gamma\sim 0$ so that $M_5$ is effectively the same as in the
DGP(+) case.

What is the nature of the beginning of the universe in GBIG1? We
can use Eq.~(\ref{in}) in Eq.~(\ref{Ray}), for matter with $w>-1$,
to analyze the initial state, $d\mu/d(h^2)\to 0+$. We find that
$h'_{\rm i}=-\infty$, i.e., infinite deceleration,
\begin{equation}
a''_{\rm i}=-\infty\,.
\end{equation}
(For GBIG2, with $d\mu/d(h^2)\to 0-$, we have $h'_{\rm
i}=+\infty$.) The initial state has no infinite-temperature big
bang, but it has infinite deceleration, and thus infinite Ricci
curvature. The brane universe is born in a ``quiescent"
singularity. (Although similar singularities may be found in
Induced Gravity models~\cite{Shtanov:2002ek}, they arise from the
special extra effect of a bulk black hole or a negative brane
tension.) The key point is that neither the DGP(+) model nor the
GB model avoid the standard big bang, as shown in
Eqs.~(\ref{bbdgp}) and (\ref{bbgb}). But together, the IG and GB
effects combine in a ``nonlinear" way to produce entirely new
behaviour. If we switch off either of these effects, the
infinite-temperature big bang reappears.

This singularity is reminiscent of the ``sudden" (future)
singularities in General Relativity~\cite{Barrow:2004xh} -- but
unlike those singularities, the GBIG1--2 singularity has finite
pressure. The initial curvature singularity signals a breakdown of
the brane spacetime. The (Minkowski) bulk remains regular, but the
imbedding of the brane becomes singular. Higher-order
quantum-gravity effects will be needed to cure this singularity.
The fact that the matter is regular at the singularity indicates
that the singularity is weaker than a standard big bang
singularity, and may be easier to ``cure" with quantum
corrections.

By performing an expansion near the initial state, using
Eq.~(\ref{fnew}), we find that the primordial Hubble rate in
GBIG1, after the infinite deceleration at the birth of the
universe, is given by
\begin{equation}
H^2\approx H_{\rm i}^2-\,H_{\rm i} \left[ {2\kappa_5^2\over
3\sqrt{r^2-32\alpha} } \right]^{1/2}(\rho_{\rm i}-\rho)^{1/2}.
\end{equation}
This is independent of the equation of state $w$. If there is
primordial inflation in the GBIG1 universe, then the acceleration
$a''$ will become positive. For a realistic model (satisfying
nucleosynthesis and other constraints), $a''$ must subsequently
become negative again, so that the universe decelerates during
radiation- and early matter-domination. Finally, $a''$ will become
positive again as the late universe self-accelerates.

\begin{figure}
\begin{center}
\includegraphics[height=3in,width=2.75in,angle=270]{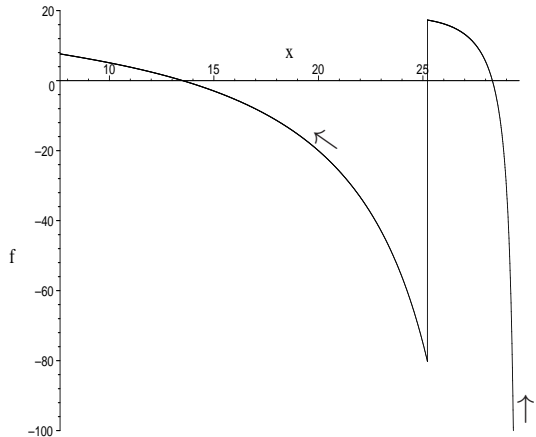}
\rput{325}(-3.45,-2.45){\large $\leftarrow$}
\rput{270}(-0.8,-6){\large $\leftarrow$} \caption{The acceleration
$f=a''/a$ vs $x=h^2$, for a GBIG1 cosmology with inflation,
followed by radiation domination, followed by late-time
self-acceleration. Arrows indicate the flow of brane proper time.
Here $\gamma=0.05$, and $n=0.8$ in Eq.~(\ref{eos}).  } \label{cos}
\end{center}
\end{figure}

We can simplify the expression~(\ref{acc}) for the acceleration in
GBIG1 via Eq.~(\ref{fnew}),
\begin{equation}
f=\frac{2x(3\gamma x+1-\sqrt{x}) +3(1+w)[x\sqrt{x}-2{x}(\gamma
x+1)]}{2(3\gamma x+1-\sqrt{x})}\!,
\end{equation}
where $f\equiv a''/a, x\equiv h^2$. For a given $w(x)$, we can
plot $f(x)$. We show an example in Fig.~\ref{cos} of a simple
model, with primordial inflation followed by radiation domination,
followed by late-time self-acceleration. We have used the
effective equation of state
\begin{equation}\label{eos}
w= \left\{ \begin{array}{lll} -0.9 && n(h_{\rm i}^2 -
h_{\infty}^2) + h_{\infty}^2< x<h_{\rm i}^2\,,\\1/3 &&
h_{\infty}^2 < x < n(h_{\rm
i}^2 - h_{\infty}^2) + h_{\infty}^2\,.\\
\end{array} \right.
\end{equation}
Here $0<n<1$ is a parameter determining the time of reheating
(with $n=0$ corresponding to no inflation and $n=1$ to no
reheating/ radiation).\\

\section{Conclusions}

In summary, the GBIG1 model provides an intriguing generalization
of the DGP(+) model -- the Gauss-Bonnet (ultra-violet) correction
to the (infra-red) Induced Gravity preserves the late-time
self-acceleration of the universe, but leads to striking new
behaviour in the early universe. Although there is still a
curvature singularity at the beginning, the density, pressure and
temperature are finite. This model deserves further investigation
as a viable cosmological model. Future work will impose
constraints on the model parameters from nucleosynthesis and
Supernova redshifts (compare Ref.~\cite{Cai:2005ie}). We expect
that these constraints will not differ appreciably from the DGP(+)
case, given the very small value of the GB parameter $\gamma$ that
is imposed by the UV-IR bootstrap. However, a nonzero $\gamma$, no
matter how small, leads to dramatic and nonperturbative changes at
high energies in the primordial universe.

From a theoretical viewpoint, it will also be important to
investigate how the GB term affects the issues of strong coupling
and ghosts in the DGP(+) cosmological model~\cite{prob}. Can the
GB term provide a lowest-order ultra-violet completion of the
DGP(+) theory? The analysis of perturbations about a Minkowski
brane with GB and IG terms~\cite{Brax:2004np} gives a starting
point for tackling the Friedmann brane case.

\[ \]{\bf Acknowledgements:} RB and RM are supported by PPARC, EP and
VZ by the NTUA research program Protagoras and by (EPEAEK
II)-Pythagoras (co-funded by the European Social Fund and National
Resources). We thank Mariam Bouhmadi-Lopez, Georgios Kofinas,
Kazuya Koyama and Alexey Toporensky for very helpful discussions.



\begin{thebibliography}{99}



\bibitem{rs}
See, e.g.,
  P.~Brax, C.~van de Bruck and A.~C.~Davis,
  Rept.\ Prog.\ Phys.\  {\bf 67}, 2183 (2004)
  [arXiv:hep-th/0404011].

\bibitem{Gregory:2000jc}
See, e.g.,
  R.~Gregory, V.~A.~Rubakov and S.~M.~Sibiryakov,
  Phys.\ Rev.\ Lett.\  {\bf 84}, 5928 (2000)
  [arXiv:hep-th/0002072];
  I.~I.~Kogan,
  arXiv:astro-ph/0108220;
  A.~Papazoglou,
  arXiv:hep-ph/0112159;
  A.~Padilla,
  Class.\ Quant.\ Grav.\  {\bf 22}, 681 (2005)
  [arXiv:hep-th/0406157];
  A.~Padilla,
  Class.\ Quant.\ Grav.\  {\bf 22}, 1087 (2005)
  [arXiv:hep-th/0410033].
\bibitem{Dvali:2000hr}
  G.~R.~Dvali, G.~Gabadadze and M.~Porrati,
  Phys.\ Lett.\ B {\bf 485}, 208 (2000)
  [arXiv:hep-th/0005016].

\bibitem{dgp0}
  H.~Collins and B.~Holdom,
  Phys.\ Rev.\ D {\bf 62}, 105009 (2000)
  [arXiv:hep-ph/0003173];
  Y.~V.~Shtanov,
  arXiv:hep-th/0005193.

\bibitem{Deffayet:2000uy}
  C.~Deffayet,
  Phys.\ Lett.\ B {\bf 502}, 199 (2001)
  [arXiv:hep-th/0010186].

\bibitem{gb}
See, e.g.,
  C.~Charmousis and J.~F.~Dufaux,
  Class.\ Quant.\ Grav.\  {\bf 19}, 4671 (2002)
  [arXiv:hep-th/0202107];
  S.~Nojiri, S.~D.~Odintsov and S.~Ogushi,
  Int.\ J.\ Mod.\ Phys.\ A {\bf 17}, 4809 (2002)
  [arXiv:hep-th/0205187];
  S.~C.~Davis,
  Phys.\ Rev.\ D {\bf 67}, 024030 (2003)
  [arXiv:hep-th/0208205];
  E.~Gravanis and S.~Willison,
  Phys.\ Lett.\ B {\bf 562}, 118 (2003)
  [arXiv:hep-th/0209076];
  J.~E.~Lidsey and N.~J.~Nunes,
  Phys.\ Rev.\ D {\bf 67}, 103510 (2003)
  [arXiv:astro-ph/0303168];
  K.~i.~Maeda and T.~Torii,
  Phys.\ Rev.\ D {\bf 69}, 024002 (2004)
  [arXiv:hep-th/0309152];
  J.~F.~Dufaux, J.~E.~Lidsey, R.~Maartens and M.~Sami,
  Phys.\ Rev.\ D {\bf 70}, 083525 (2004)
  [arXiv:hep-th/0404161];
  T.~G.~Rizzo,
  JHEP {\bf 0501}, 028 (2005)
  [arXiv:hep-ph/0412087].

\bibitem{Mavromatos:2005yh}
  N.~E.~Mavromatos and E.~Papantonopoulos,
  arXiv:hep-th/0503243.

\bibitem{alpha}
The assumption that $\alpha$ is non-negative is motivated by
string theory, where typically $\alpha\propto +L_{\rm string}^2$.

\bibitem{Kofinas:2003rz}
  G.~Kofinas, R.~Maartens and E.~Papantonopoulos,
  JHEP {\bf 0310}, 066 (2003)
  [arXiv:hep-th/0307138].

\bibitem{ads}
Note that the $\Phi=-1/2\alpha$ branch corresponds to an anti de
Sitter bulk, even though there is no cosmological constant.

\bibitem{dgppert}
  T.~Tanaka,
  Phys.\ Rev.\ D {\bf 69}, 024001 (2004)
  [arXiv:gr-qc/0305031];
  C.~Deffayet,
  Phys.\ Rev.\ D {\bf 71}, 103501 (2005)
  [arXiv:gr-qc/0412114];
  K.~Koyama and K.~Koyama,
  Phys.\ Rev.\ D {\bf 72} (2005) 043511
  [arXiv:hep-th/0501232].

\bibitem{bmp}
M. Bouhmadi-Lopez, unpublished notes.

\bibitem{Shtanov:2002ek}
  Y.~Shtanov and V.~Sahni,
  Class.\ Quant.\ Grav.\  {\bf 19}, L101 (2002)
  [arXiv:gr-qc/0204040].

\bibitem{Barrow:2004xh}
  J.~D.~Barrow,
  Class.\ Quant.\ Grav.\  {\bf 21}, L79 (2004)
  [arXiv:gr-qc/0403084].

\bibitem{Cai:2005ie}
  R.~G.~Cai, H.~S.~Zhang and A.~Wang,
  arXiv:hep-th/0505186.

\bibitem{prob}
  C.~Deffayet, G.~R.~Dvali, G.~Gabadadze and A.~I.~Vainshtein,
  Phys.\ Rev.\ D {\bf 65} (2002) 044026
  [arXiv:hep-th/0106001];
  M.~A.~Luty, M.~Porrati and R.~Rattazzi,
  JHEP {\bf 0309}, 029 (2003)
  [arXiv:hep-th/0303116];
  V.~A.~Rubakov,
  arXiv:hep-th/0303125;
  G.~Dvali,
  arXiv:hep-th/0402130;
  A.~Nicolis and R.~Rattazzi,
  JHEP {\bf 0406}, 059 (2004)
  [arXiv:hep-th/0404159];
  N.~Kaloper,
  Phys.\ Rev.\ D {\bf 71}, 086003 (2005)
  [Erratum -- ibid.\ D {\bf 71}, 129905 (2005)]
  [arXiv:hep-th/0502035];
  K.~Koyama,
  arXiv:hep-th/0503191;
  C.~Deffayet and J.~W.~Rombouts,
  arXiv:gr-qc/0505134.

\bibitem{Brax:2004np}
  C.~Charmousis and J.~F.~Dufaux,
  Phys.\ Rev.\ D {\bf 70}, 106002 (2004)
  [arXiv:hep-th/0311267];
  P.~Brax, N.~Chatillon and D.~A.~Steer,
  Phys.\ Lett.\ B {\bf 608}, 130 (2005)
  [arXiv:hep-th/0411058].



\end{thebibliography}
\end{document}